\documentclass[letters,fleqn,usenatbib,useAMS]{mnras}

\usepackage{graphicx}	
\usepackage{amsmath}	
\usepackage{amssymb}	
\usepackage{multicol}        
\usepackage{pdflscape}	
\usepackage{tablefootnote}
\usepackage[flushleft]{threeparttable}
\usepackage{hyperref}     

\usepackage[T1]{fontenc}
\usepackage{ae,aecompl}

\usepackage{txfonts}

\title[Superhump ocurrence in the Nova RR Pic]{Photometric long-term variations and superhump occurrence in the Classical Nova RR Pictoris}

\author[I. Fuentes-Morales et al.]{I. Fuentes-Morales$^{1}$\thanks{E-mail:
irma.fuentes@uv.cl (IF), nikolaus.vogt@uv.cl (NV)
}, N. Vogt$^{1}$\footnotemark[1],  C. Tappert$^{1}$, L. Schmidtobreick$^{2}$, F.-J. Hambsch$^{3}$ \newauthor and M. Vu\v{c}kovi\'c$^{1}$   \\
\\
$^{1}$ Instituto de F\'isica y Astronom\'ia, Universidad de Valpara\'iso, Avda. Gran Breta\~na 1111, Valpara\'iso, Chile\\
$^{2}$ European Southern Observatory, Casilla 19001, Santiago 19, Chile\\
$^{3}$ Vereniging Voor Sterrenkunde (VVS), Oude Bleken 12, 2400 Mol, Belgium}

\date{Accepted XXX. Received YYY; in original form ZZZ}

\pubyear{2017}

\begin{document}
\label{firstpage}
\pagerange{\pageref{firstpage}--\pageref{lastpage}}
\maketitle

\begin{abstract}
We present an analysis of all available time-resolved photometry from the literature and new light curves obtained in 2013--2014 for the old nova RR Pictoris. The well-known hump light curve phased with the orbital period reveals significant variations over the last 42 years in shape, amplitude and other details which apparently are caused by long-term variations in the disc structure. In addition we found evidence for the presence of superhumps in 2007, with the same period ($\sim$9$\%$ longer than the orbital period), as reported earlier by other authors from observations in 2005. Possibly, superhumps arise quickly in RR Pic, but are sporadic events, because in all the other observing runs analysed no significant superhump signal was detected. We also determined an actual version of the Stolz--Schoembs relation between superhump period and orbital period, analysing separately dwarf novae, classical novae and nova-like stars, and conclude that this relation is of general validity for all superhumpers among the cataclysmic variables (CVs), in spite of small but significant differences among the sub-types mentioned above. We emphasize the importance of such a study in context with the still open question of the interrelation between the different sub-classes of CVs, crucial for our understanding of the long-term CV evolution. 

\end{abstract}




\section{Introduction}
 Cataclysmic variables (CVs) are close interacting binary systems where a late-type dwarf star fills its Roche lobe and transfers mass towards the white dwarf (WD) primary component. In case of a weak --or absent-- magnetic field of the WD, the transferred material from the secondary star is accumulated forming an accretion disc around the WD. Once the accumulated mass on the primary reaches a critical value, the surface temperature of the WD increases and triggers nuclear fusion reactions producing a thermonuclear runaway, which is known as a nova eruption. CVs that underwent a nova eruption are called classical novae (for details see \citealt{warner95}).

RR Pictoris is one of the brightest and nearest novae known (distance $d=388 \pm 88\,\mathrm{pc}$; \citealt{gaiaDist}). Its nova eruption was discovered by \citet{jones28} at its maximum brightness $V=1.2$ mag in 1925. Currently the stellar remnant appears to be in a state of quiescence, with  $V \sim 12.7$ mag. The photometric light curves of RR Pic show as a permanent feature a periodic hump of $P_\mathrm{orb} =3.48$\,h, which has been identified as the orbital one \citep{vogt75}. Additionally, they have undergone several changes and transient features over the decades since its eruption, showing eclipse-like characteristics \citep{warner86}, 15\,min periodicities \citep{kubiak84}, QPOs and positive superhumps \citep{linda2008}. Time-series spectroscopy and Doppler mapping found evidence for two emission sources, the hotspot and a source opposite. \citep{haefner82,linda2003,ribeiro2006}. In a recent work \citet{vogt2016} derived a more precise orbital ephemeris from all available photometric humps observed in the past five decades. They furthermore found variations in the O--C diagram that could imply the presence of a third body.\\
\indent
The motivation for the present investigation was the discovery already mentioned of an additional period that is 8.6 per cent longer than the orbital one, which was interpreted as a superhump period ($P_\mathrm{sh}$) by \citet{linda2008}.
Superhumps are photometric modulations in the light curve with periods slightly different from  the orbital period, which are present in SU-UMa type dwarf novae during superoutburst \citep{vogt74} and also as `permanent superhumps' in some nova-like CVs \citep{patterson99}. If $P_\mathrm{sh}$ is a few percent longer than the orbital period we call this a ``positive superhump'', while the case $P_\mathrm{sh}$ < $P_\mathrm{orb}$ refers to ``negative superhumps''. \citet{vogt82a}  was the first who proposed superhumps are associated to an eccentric accretion disc during superoutburst, based on spectroscopic observations of Z Cha. Later studies have established that the disc becomes eccentric when it reaches the 3:1 resonance, due to the interaction with the secondary, causing the elongated section of the disc moves to different time-scale respect to the orbital period (\citealt{whitehurst88,osaki89,patterson2005} and references therein). Thus, the presence of superhump in classical novae can be used as a tool to understand the state of the system in general, and of the disc properties decades after the nova eruption. \\

\noindent
In this paper, we present a re-analysis of all available photometric data of RR Pic, between 47 and 89 years after its classical nova explosion observations. We also add new data taken during 2013--2014 and study the orbital hump light curves during the past decades in a systematic way. Furthermore we perform a methodical search for superhumps in RR Pic and compare our results with those known for other classical novae, nova-like stars and dwarf novae. 
  
\section[]{Observations}
\subsection{Compilation of historical light curves and new observations}

The light curves of RR Pic analysed here consist of published observations made by different authors, complemented by new data obtained in 2013 and 2014. The different campaigns are summarised in Table \ref{tab:t1}, which contains: the Heliocentric Julian date (HJD) at start and at the end of each campaign ($\rmn{HJD}_{\rmn{start}}$ and $\rmn{HJD}_{\rmn{end}}$), the total number of light curves (N), the total time covered by observations ($\Delta{t}$), the identification of the observer and the campaign name (see Fig. \ref{fig:po-all}). An observing log for hitherto unpublished data is given as on-line supplementary data. In particular, we can distinguish the following campaigns: 

\begin{enumerate}

\item {\it UBV\/} photometry observed by \citet{vogt75} during four nights in December 1972 and three nights of white light data in January 1974; from the {\it UBV\/} data we analysed only  the {\it V\/} band light curves, given that the color index {\it B -- V \/} is close to 0 (see fig. 5 in \citealt{vogt75}). All these observations have covered more than one orbital cycle.

\item High-speed photometry in white light of \citet{warner86} from 1972 to 1984 over 23 nights. In eight of them at least one orbital period was covered. Only raw instrumental values (count rates per second) are available, without observations of comparison stars nor any other calibration. Details on how this data were analysed are described in section~\ref{subsect:homog}.

\item Light curves in five nights observed in January 1980 by \citet{haefner82} in the {\it V\/} band. In all of them at least one complete orbital cycle was observed.

\item \citet{kubiak84} obtained {\it UBV\/} photoelectric photometry in five nights in October 1982. We analysed the published {\it B\/} differential band light curves. Only in one night an entire orbital period was covered.

\item The Center Backyard Astrophysics (CBA)\footnote{http://cbastro.org/} kindly provided us with additional unpublished differential photometry in white light of RR Pic, obtained in a total of 50 nights between 1999 and 2007 by different amateur astronomers. 

\item Differential photometric data obtained by \citet{linda2008} during the first semester of 2005, covering 18 complete orbital cycles.
\end{enumerate}
\indent
In addition one of us (FJH) has obtained new {\it V\/} band photometry in 81 nights between 21 November 2013 and 16 March 2014, using the robotic 40 cm telescope at the Remote Observatory Atacama Desert (ROAD), situated in San Pedro de Atacama, Chile \citep{hambshroad2012}. This telescope is equipped with a CCD camera that contains a Kodak 16803 chip of $4\rmn{k} \times 4\rmn{k}$ pixels of 9-$\,\umu$m size. The integration time was 30 seconds with a cadence of 2--3 minutes. Normally, one entire orbital period per night was covered. However, due to the coordinates of RR Pic, only a partial coverage of the orbital period was possible at the end of this run (February to March 2014).

We had access to the original data of \citet{vogt75}, \citet{warner86}, \citet{linda2008} and those of the Center Backyard Astrophysics, all in digital form. The light curves of \citet{haefner82} and \citet{kubiak84} were only available as figures. Therefore, we digitised them from the respective publications, using the \textsc{Dexter}\footnote{http://dc.zah.uni-heidelberg.de/dexter/} service from the website of the German Astrophysical Virtual Observatory (GAVO).

\begin{figure}
  \centering  
  \resizebox{0.485\textwidth}{!}{\includegraphics[trim={0mm 33mm 0mm 0mm}, clip]{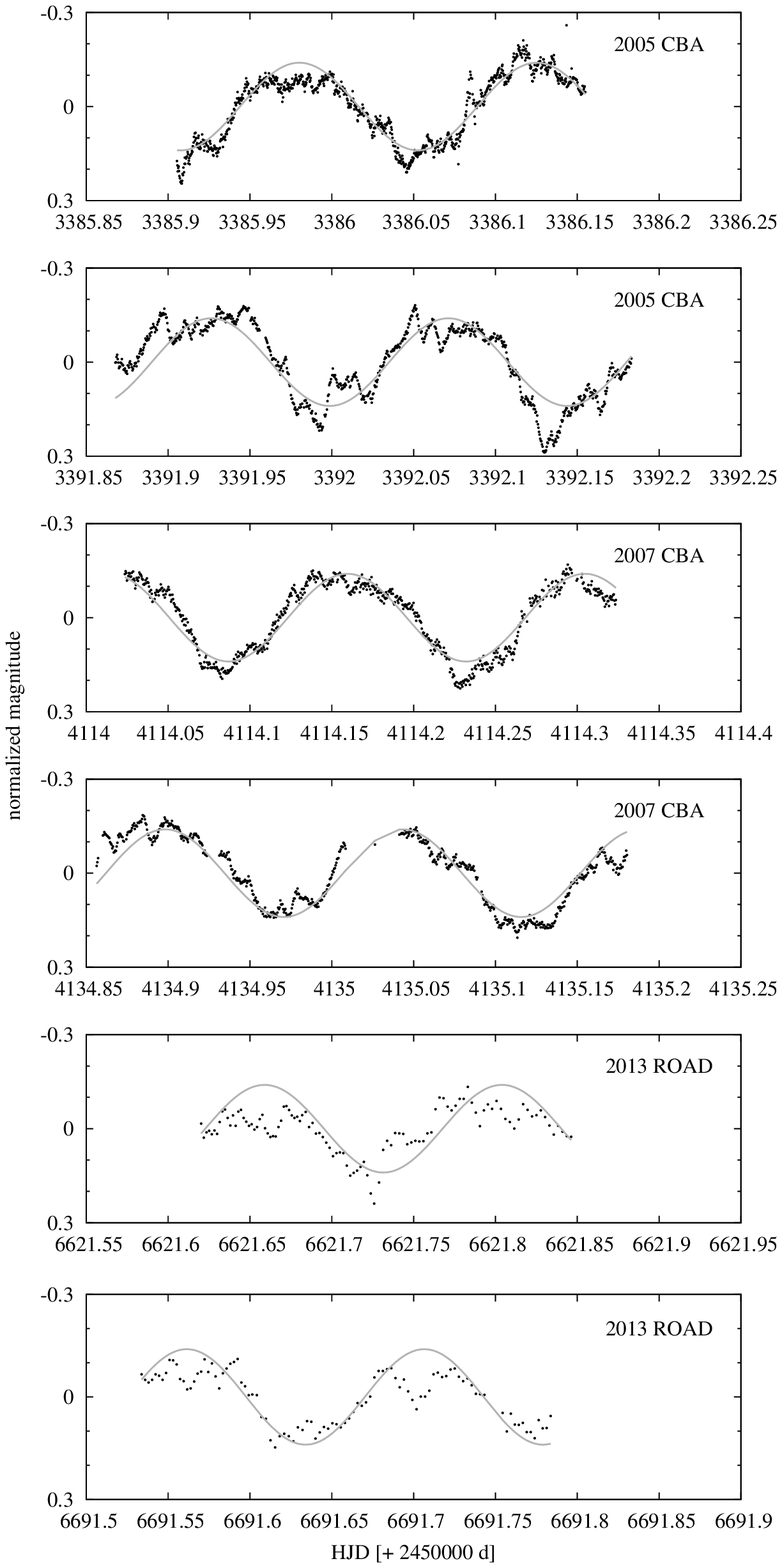}}
  
  \caption{Examples of unpublished RR Pic light curves from CBA database and from ROAD Observatory are shown as black points. The sinus fit adjusted to the light curves as detailed in the text is shown as grey line. Axis Y correspond to normalised differential magnitude and axis X to reduced HJD.}
  \label{fig:selectedLC}
\end{figure}

In Fig. \ref{fig:selectedLC} we present examples of the light curves corresponding to hitherto unpublished campaigns: 2005 and 2007 from CBA and 2013 from ROAD. In all light curves we also show the mean sine fit curve of the orbital hump with an adopted amplitude $A=0.14\pm0.01$ mag and the global orbital period according to ephemeris (\ref{eq:ef}). 

\begin{table}
\begin{minipage}{1.0\columnwidth}

 \caption{Summary of RR Pic photometric data for each campaign.}

\setlength{\tabcolsep}{0.17cm}
\begin{tabular}{@{}llllll}

\hline\noalign{\smallskip}
  $\rmn{HJD}_{\rmn{start}}$  & $\rmn{HJD}_{\rmn{end}}$   & N  & $\Delta{t}$  & Observer & Campaign   \\
 (+2400000 d)          & (+2400000 d)                 &         &   (hr)     &       &   name\\
\hline
41656.40332            & 41696.30201         & 11     &  36.6       &  1, 4  &  1972--1973\\  
42023.52854            & 42064.69775         &  7     &  18.8       &  1, 4  &  1973--1974\\
44255.55140            & 44261.55750         &  5     &  31.1       &   2    &  1980--1 \\
4294.267740            & 4295.267660         &  2     &   6.8       &   4    &  1980--2 \\
44937.36686            & 44951.46229         &  5     &  15.3       &   4    &  1981\\
45245.77872            & 45253.71943         &  4     &  12.8       &  3     &  1982\\
51875.98085            & 51967.91066         & 15     &  84.9      & 5       &  2000--2001\\
53375.02367            & 53398.87515         & 15     &  87.7      & 5       &  2005--1\\
53409.53241            & 53470.47910         & 14     &  66.3      & 6       &  2005--2\\
54114.02361            & 54138.98391         & 6      & 45.7       & 5       &  2007  \\
56617.63134            & 56732.50956         & 81     & 449.05     &   7     &  2013--2014\\
\hline\noalign{\smallskip}
                        
\end{tabular}
\\

1. \citet{vogt75}, 2. \citet{haefner82}, 3. \citet{kubiak84}, 4. \citet{warner86}, 5. CBA database, 6. \citet{linda2008}, 7. ROAD

\label{tab:t1}

\end{minipage}

\end{table}

\subsection{Homogenization of the data}
\label{subsect:homog}
All historical data were converted to a standard format: HJD and differential magnitude with respect to the comparison star used by \citet{vogt75}. Additionally the data were normalised by subtracting the mean magnitude of each data set. \citeauthor{warner86}'s uncalibrated data files were treated in a special way, transforming his counts per second into magnitudes and adjusting their zero points to fit \citeauthor{vogt75}'s phased light curves. This was possible because both data sets cover partly the same epoch. For the same reason, we combined \citeauthor{vogt75}'s and \citeauthor{warner86}'s data in two campaigns, 1972--1973 and 1973--1974. The 1972--1973 campaign contains all {\it B\/} band light curves obtained during 1972 by both authors, and observations from 1973 January 11 and 13 from \citeauthor{warner86}. The 1973--1974 campaign was composed of white light data from seven nights between 1973 December 7 and 1974 January 17 from both authors. \citeauthor{haefner82}'s light curves (campaign 1980--1) were combined with two light curves of \citeauthor{warner86} (campaign 1980--2), which were observed in high resolution compared with \citeauthor{haefner82}. To obtain the same time resolution as in the 1980--1 campaign, we re-binned the data set of \cite{warner86}.

\section{Results}

\subsection{Variations of orbital light curve}

\begin{figure}
  \resizebox{0.459\textwidth}{!}{\includegraphics[trim={0mm 17mm 0mm 0mm}, clip]{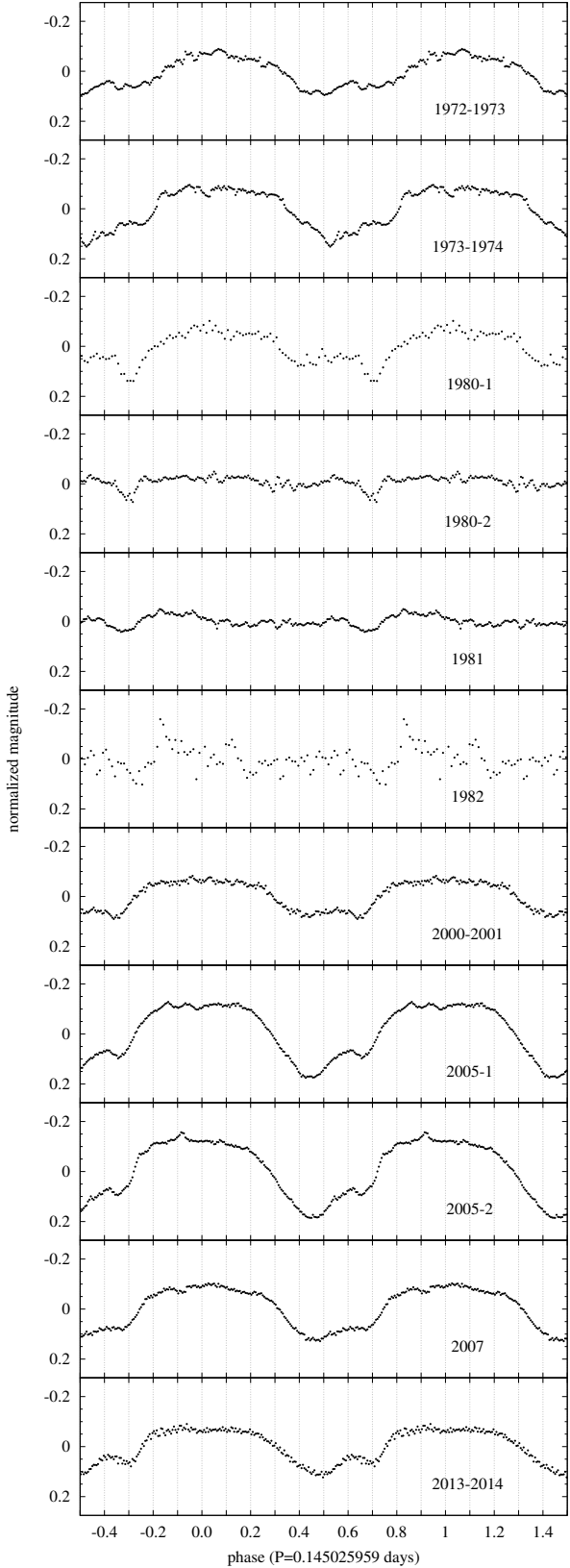}}
 \caption{Orbital phase of the classical nova RR Pic over the last 42 years, folded with the ephemeris given in Eq. (\ref{eq:ef}). Variations of the characteristic hump feature are clearly visible over time.  }
  \label{fig:po-all}
\end{figure}

\noindent
We used the \textsc{period04} software \citep{period04} to search for peridiocities in the light curve via Discrete Fourier Transform (DFT).  All campaigns are sampled until the Nyquist frequency. However, since we are searching for frequencies close to the orbital one, the examined frequency range was from 0 to 30 c/d. For all campaigns the respective periodogram shows the known orbital frequency as the dominant peak close to $f_{0}=6.895$ c/d, corresponding to $\rmn{P_{orb}}=3.48~ \rmn{h}$. The orbital phase of each sample was calculated using the new long-term ephemeris of the orbital hump maximum: 
\begin{equation}
 {\rmn{HJD(max)}}=2438815.3664(15) + 0^{{\rmn{d}}}.145025959(15){\rmn{E}}
 \label{eq:ef}
\end{equation}
determined by \citet{vogt2016}. The mean phased-folded light curves were computed by averaging the data points into 0.02 phase bins. The results are shown in Fig.~\ref{fig:po-all}. 

The orbital light curve has notably changed its shape over the course of the last 42 years (Fig.~\ref{fig:po-all}). During the data from 1972 to 1974 the light curves predominantly showed humps of amplitude $\sim$0.2$^\rmn{m}$ and smooth minima at phase 0.55, being about 0.05$^\rmn{m}$ deeper in 1973--1974 than one year before. In addition, there was a step-like feature at phase 0.65--0.70. From 1980 to 1982, the light curve underwent dramatic changes: in January 1980, the orbital hump was still present, although with a smaller amplitude, and there was a deep minimum at phase 0.7 (1980--1). However, in two observing runs obtained only one month apart (1980--2), the orbital hump practically disappeared, showing a predominantly flat light curve, interrupted only by the minimum at phase 0.7. Then in 1981, there was a very shallow maximum at phase $\sim$0.9, immediately after the minimum around phase 0.7 which now seems to be broader than before. A similar light curve was observed in 1982 with a larger scatter, possibly due to enhanced flickering. The earlier step-like feature already present in the 1970s at phase $\sim$0.7 has transformed into a relatively sharp eclipse-like feature in 1980--1982, seen at the same phase which represents, at this epoch, the total minimum of the light curve. Almost twenty years later, in 2000--2001 the orbital hump had recovered with total amplitude of $\sim$0.15$^\rmn{m}$ followed by a flat minimum at phases 0.40--0.65. In 2005 the hump amplitude had increased considerably to about 0.28$^\rmn{m}$, showing now a pronounced minimum at phase 0.45 and a small secondary minimum at phase 0.7. In 2007 the hump amplitude had diminished to 0.20$^\rmn{m}$, and the step-like feature at phase 0.7 was recovered, similar to the situation in the 1972 to 1974. More recently, in 2013--2014 we observe still slightly reduced hump amplitudes (0.18$^\rmn{m}$) with a pronounced main minimum at phase 0.5 and a shallow secondary one at phase 0.7.

We can summarise the orbital light curve properties of RR Pic during the past 42 years in the following way: the dominant orbital hump was always present, but with variable amplitude between 0.28$^\rmn{m}$ in 2005 and $\sim$0.05$^\rmn{m}$ in the early 1980s. As a second permanent light curve property we found a secondary minimum or step-like feature at phases 0.65--0.7 which seems to be always present, but more pronounced as an `eclipse' corresponding to the total minimum in the whole orbital light curve, whenever the hump amplitude is low, or less pronounced (`step-like feature') when the orbital hump had a large amplitude ($>$~0.2$^\rmn{m}$).

\subsection{Search for superhumps: a new detection in 2007}

\noindent
In RR Pic, the superhump phenomenon is an additional periodic modulation in the light curve with smaller amplitude than that of the orbital hump. Therefore, a search for superhumps is only possible in the residual data after subtraction of the dominant frequency. Any additional frequency found close to the orbital one could be potentially associated with a $P_\mathrm{sh}$. For sufficiently long data sets, the presence of the beat period of about 1.8 days provides an additional confirmation. 

\begin{figure}
  \centering  
   \resizebox{0.5\textwidth}{!}{\includegraphics[trim={0mm 0mm 0mm 28mm}, clip]{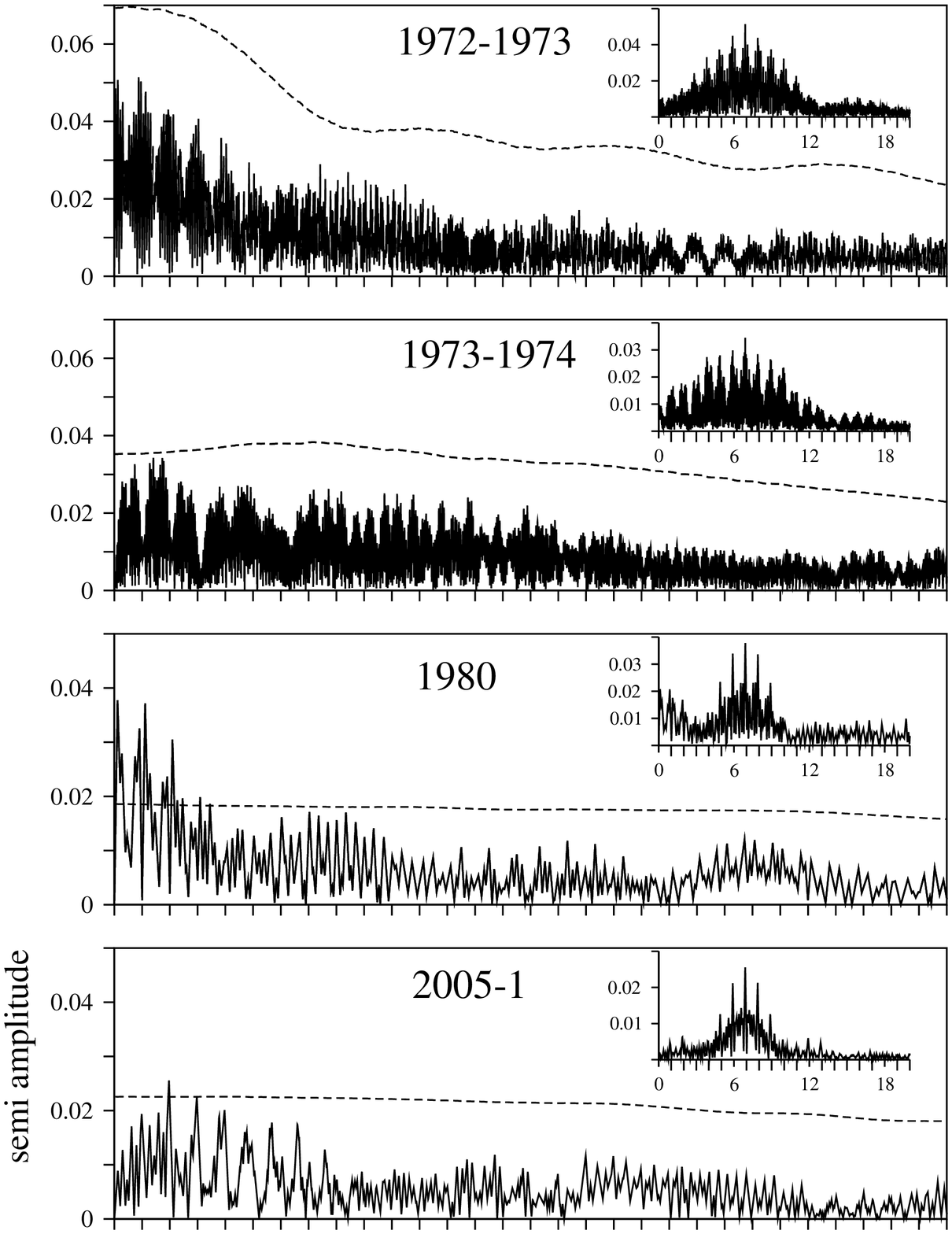}}
    \resizebox{0.5\textwidth}{!}{\includegraphics[trim={0mm 0mm 0mm 77.6mm}, clip]{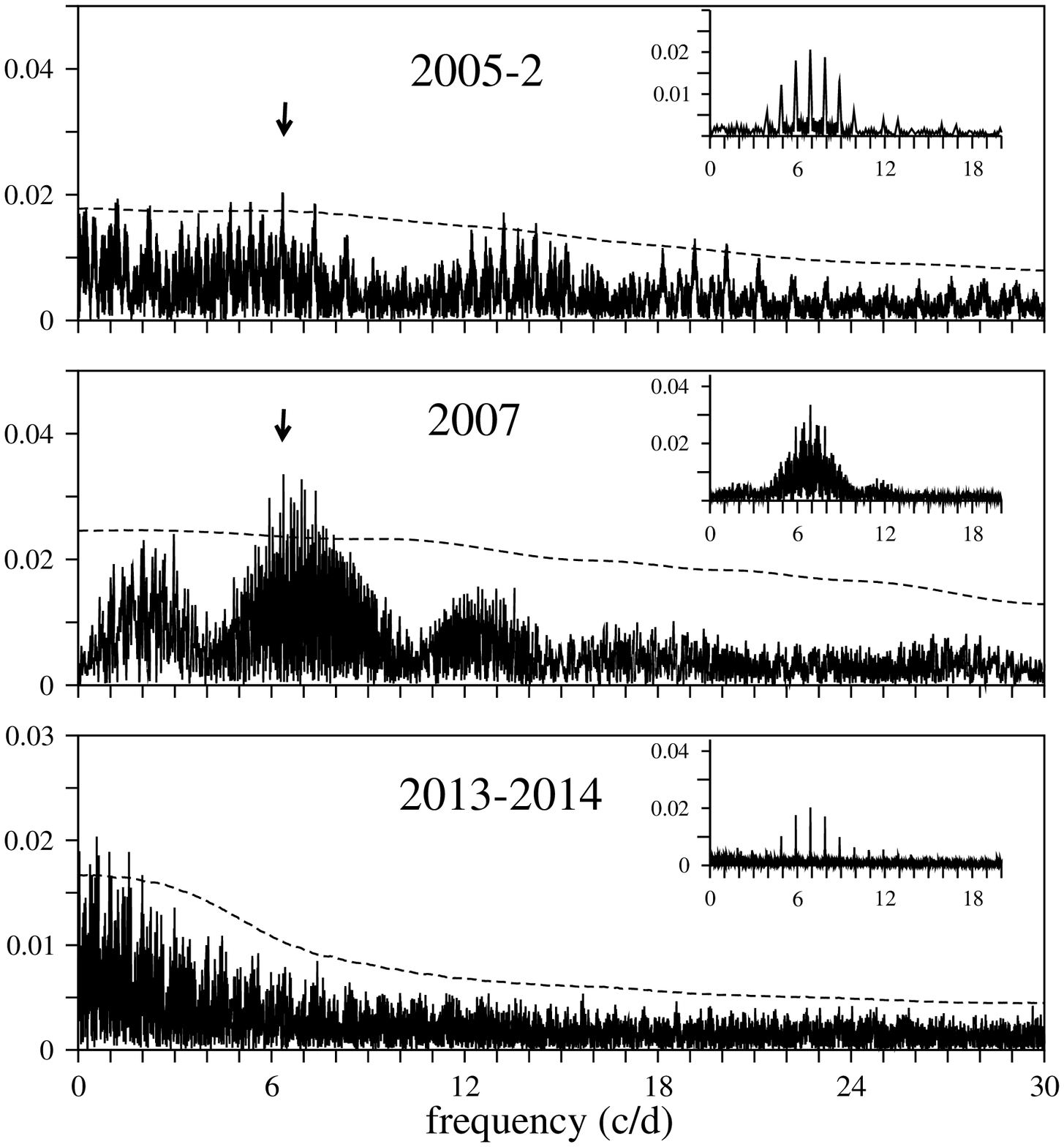}}
 \caption{Periodigrams for the residual data after subtracting the orbital frequency and its harmonics for selected campaigns (see text). The dashed line corresponds to the noise level of 3$\sigma$, the arrows point to suspected superhump frequencies and the inset plot shows the spectral window centred at the dominant orbital frequency.}
  \label{fig:periodig-SH}
\end{figure}

To ensure that the residual light curves of all campaigns were not contaminated by the presence of harmonics and aliases of the orbital period, we extracted up to the sixth orbital harmonics from the data. In order to check that the orbital features have been completely removed we have folded these residuals to the phase of $P_\mathrm{orb}$. These folded light curves were completely flat.
In this manner, we analyse all campaigns. In Fig.~\ref{fig:periodig-SH} we show the Fourier spectra of the residuals light curves for selected campaigns which covered at least two orbital cycles.
To check the possible presence of any periodic signal associated directly to the sampling, we created an artificial sample containing the same time values of each set data versus a constant magnitude value. Periodograms of those artificial data represent the individual spectral windows, which are shown as inserts in Fig.~\ref{fig:periodig-SH}, centred on the orbital frequency and scaled to the residual amplitudes for each campaign. The noise spectrum was calculated as the average of the remainder amplitudes in certain frequency steps of a size defined by the box size parameter (see \citealt{period04} for details) in the range from 0 to 30 c/d. A signal near the orbital period that exceeds the 3$\sigma$ noise level, which corresponds to 80\% confidence limit for non-normal distributions (\citealt{kuschnig97noiselevel}, dashed line in Fig.~\ref{fig:periodig-SH}), is considered for further superhump analysis. Only in the 2005 data \citep[][hereafter campaign 2005--2]{linda2008} and the 2007 CBA campaign we found signals according to these criteria, which are highlighted with an arrow in Fig.~\ref{fig:periodig-SH}.

\begin{figure}
  \centering  
  \resizebox{0.5\textwidth}{!}{\includegraphics{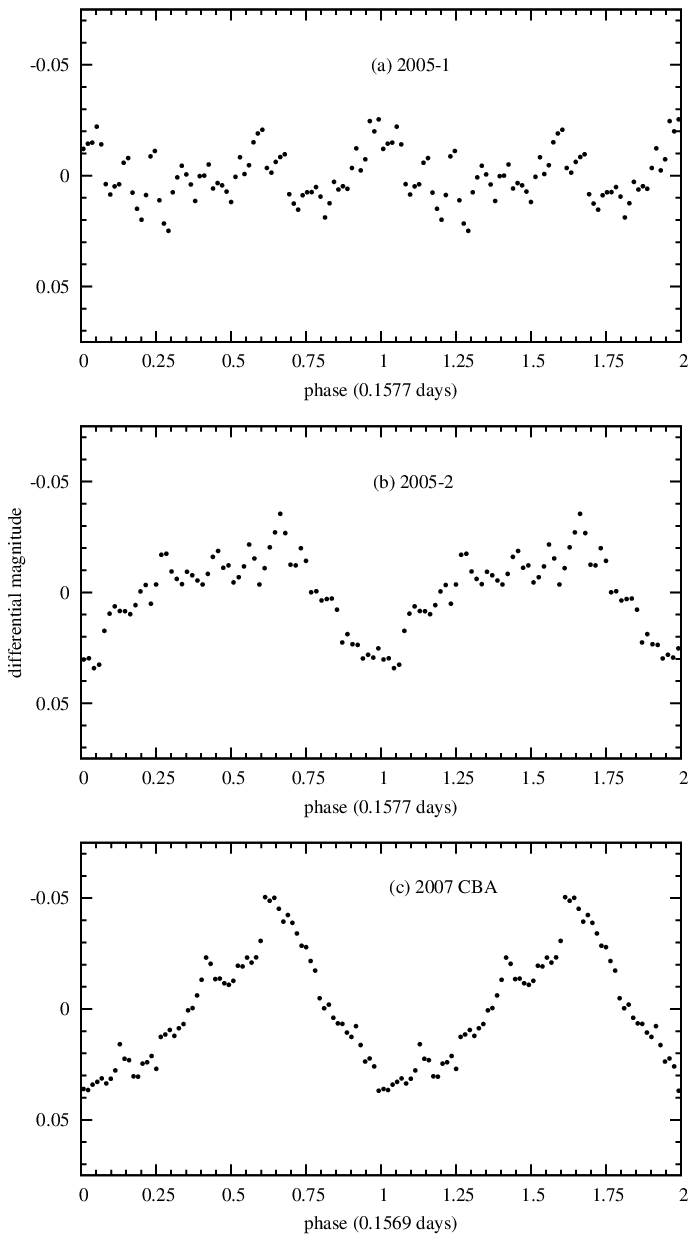}}
 \caption{Superhump phase averaged into 0.02 phase bins for normalised differential magnitude light curves after subtracting the orbital frequency. \textbf{(a)} residuals of the 2005--1 data folded with superhump ephemeris (\ref{eq:sh2005}) \textbf{(b)} the same but for the 2005--2 data \textbf{(c)} 2007 data folded with superhump ephemeris (\ref{eq:sh2007}).}
  \label{fig:SH}
\end{figure}

Although the 2005 CBA observations (campaign 2005--1) were performed only about one month before the 2005--2 campaign (see Table~\ref{tab:t1}), we did not find any significant signal close to the 2005--2 signal ($f_\rmn{sh}=6.33 \pm 0.02$\,c/d; $P_\rmn{sh}=3.79 \pm 0.01$\,h, where we have used the actual frequency resolution of this campaign as error in frequency, i.e., $1/\Delta t$). In order to check whether this null detection can be attributed to the time resolution and the coverage of the data instead of the real absence of the superhump, we folded those with ephemeris corresponding to the first minimum observed in 2005--2.
\begin{equation}
 {\rmn{HJD(min)}}=2453409.54719 + 0^{{\rmn{d}}}.1577(4)~{\rmn{E}}
\label{eq:sh2005}
\end{equation}
Comparing both averaged phased light curves, the superhump found in 2005--2 (February--April; Fig.~\ref{fig:SH} b) is clearly absent in the 2005--1 data (January; Fig.~\ref{fig:SH} a). 
Instead, we do find a clear detection of the superhump in the 2007 data. This campaign was composed of six light curves observed between January 13 and February 7 in 2007. The residual spectrum showed several peaks close to the orbital signal. The highest peak found at $f_\rmn{sh}=6.37 \pm 0.04$\,c/d is equivalent to $P_\mathrm{sh}=3.77 \pm 0.02$\,h. The other neighbouring peaks refer to one-cycle per day aliases, as is evident from the spectral window (Fig. \ref{fig:periodig-SH}). Choosing the minimum as zero point for the superhump phase yields the ephemeris                                                                 
\begin{equation}
 {\rmn{HJD(min)}}=2454114.07167 + 0^{{\rmn{d}}}.1569(6)~{\rmn{E}}
\label{eq:sh2007}
\end{equation}
 \noindent
 The corresponding phase-folded light curve is shown in Fig. \ref{fig:SH} (c). Within the errors, this new superhump period is identical to that found by \cite{linda2008} in 2005, with a period excess $\epsilon \approx 8.6$ per cent longer than the orbital period ($\epsilon=(P_\mathrm{sh}-\mathrm{P_{orb}})/\mathrm{P_{orb}}$, \citealt{patterson98}). The amplitudes of the superhump were $\sim \rmn{0.07^{m}}$ and $\sim \rmn{0.09^{m}}$ in 2005--2 and 2007, respectively, corresponding to 25 and 45 per cent of the orbital hump amplitude. 
The superhump light curves shapes in both sets are rather similar, being slightly asymmetric with maxima around superhump phases 0.6--0.7 (Fig. \ref{fig:SH} b and c).
 

\section{Discussion}

\subsection{Is RR Pic a permanent or a sporadic superhumper?}

The extra periodicity $P_\mathrm{sh}=3.77\pm0.02$ h found in 2007 data, which is comparable to the detection found in 2005 ($P_\mathrm{sh}=3.79\pm0.01$\,h), certainly supports the idea that RR Pic is an old nova with a reformed disc sufficiently hot and stable to present positive superhumps.

However, only in two out of eleven comparatively short-spanned data sets taken in the last 42 years, superhumps could be detected. RR Pic is not a permanent superhumper. Especially striking is the emergence of superhumps within the 10 days that mark the time between the 2005--1 and the 2005--2 data set. They are also present in the next available data set, 2007. However, we have no information on the behaviour during the 643 days in-between. Were the superhumps present all through that time? Our newest data set, 2013--2014, has been obtained more than 6 years later and cover several months, but no extra frequency up to 3$\sigma$ level close to the orbital frequency was found. The superhumps of 2005 and 2007 had disappeared.

\begin{figure}
  \centering  
  \resizebox{0.48\textwidth}{!}{\includegraphics{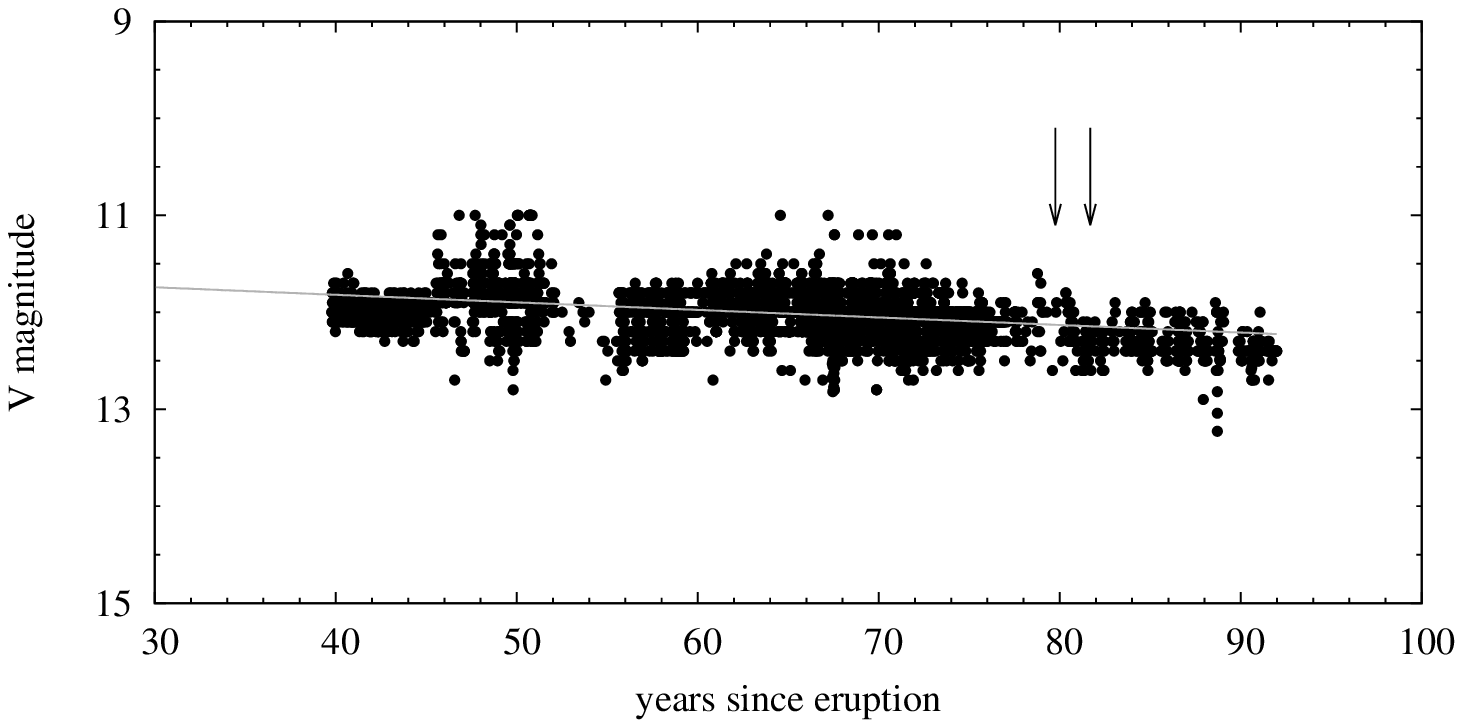}}

 \caption{{\it AAVSO} light curve of RR Pic from 1965 until May 2017. A slight decline in brightness is observed during the last five decades. Epochs where superhumps were detected are marked with arrows in the plot.}
  \label{fig:aavso}
\end{figure}

Our current understanding of the superhump phenomenon is that it is correlated to the relative dimension of the accretion disc \citep{,whitehurst88,patterson2005}. Assuming that a larger disc is also more luminous, the superhump in RR Pic should be accompanied by a corresponding change in the brightness of the system. Nevertheless, observations recorded in the American Association of Variable Star Observers ({\it AAVSO}\footnote{https://www.aavso.org/lcg}) do not show a significant variation in the brightness during these epochs (see Fig. \ref{fig:aavso}). Instead, the light curve presents a comparatively smooth decrease with a slope of $+$7.8$\pm$0.2\,mmag/yr during the last five decades, which indicates that RR Pic has not yet reached a stable state. Furthermore, the orbital phased light curves in epochs with superhumps (see in Fig. \ref{fig:po-all} campaigns 2005--1, 2005--2 and 2007) did not exhibit any eclipse feature although the disc is expected to be larger then normally. 

RR Pic is catalogued as a SW Sex star \citep{linda2003}. This subtype is characterised by high mass transfer rates, which could result in a disc sufficiently hot and stable to reach the 3:1 resonance, where the tidal forces cause it to become elliptical. Since the superhump seems to be a sporadic event, the border of the disc of RR Pic is probably located near the 3:1 resonance level, implying that small variations in mass transfer rate and/or disc size trigger occasionally the appearance of superhumps, without being accompanied by significant variations in the overall disc brightness.

\begin{table*}

 \caption{Properties of post-novae and nova-like stars with confirmed positive superhumps. }

 \begin{threeparttable}
\begin{tabular}{lrrrrcc}

\hline
                       & Name              & Type & $P_{\rmn{orb}}$  & $P_{\rmn{sh}}$   & Ref.               & Remarks  \\
                       &                   &      &   (d)            &  (d)             &                    &           \\
\hline
\hline

\textbf{Classical}     &  CP Pup (1942)    &Na    & 0.06126          & 0.0625            &   (1), (2)        &        \\
\textbf{Novae}         &  V1974 Cyg        &Na    & 0.08126          & 0.08522           &  (3), (4)         &    (a) \\
                       &  V630 Sgr         &Na    & 0.11793          & 0.1242            &     (5), (6)      &        \\
                       &  V4633 Sgr        &Na    & 0.12557          & 0.1277            &     (6), (7)      &    (b) \\
                       &  V603 Aql (1918)  &Na    & 0.13820          & 0.14646           &   (8), (9)        &    (a) \\
                       &   RR Cha          &Na    & 0.14010          & 0.1444            &     (10)          &        \\
                       &   RR Pic (1925)   &Nb    & 0.14502          & 0.1577            &  (11), (12)       &        \\

\hline

\textbf{Nova-like}     &  BK Lyn           &      & 0.07498          & 0.0786            &   (13), (14)      & (a), (c)\\
\textbf{stars}         &  V348 Pup         &      &  0.10184         & 0.10857           &   (15), (16)      &        \\
                       &  V795 Her         &      &  0.10825         & 0.11649           &   (17), (18)      &        \\
                       &  J1924+4459       &      & 0.11438          & 0.12245           &   (19), (20)      &        \\ 
                       &  V592 Cas         &      & 0.11506          & 0.12228           &      (21)         &     (a)\\
                       &   LQ Peg          &      & 0.12475          & 0.143             &      (22)         &        \\
                       &   AH Men          &      &0.12721           & 0.1385	         &      (23)         &        \\ 
                       &   MV Lyr          &      &0.13233           & 0.1377            &      (24)         &        \\
                       &   DW UMa          &      & 0.13661          & 0.146             &      (25)         &     (a)\\ 
                       &   TT Ari          &      & 0.13755          & 0.14881           &      (26)         &     (a)\\ 
                       &   PX And          &      & 0.14635          & 0.1592            &      (23)         &     (a)\\
                       &   AO Psc          &      & 0.14963          & 0.1658            &      (27)         &        \\
                       &   BZ Cam          &      &  0.15353         & 0.15634           &      (28)         &        \\
                       &   BB Dor          &      &  0.15409         & 0.1632            &      (29)         &        \\
                       &   BH Lyn          &      &0.15587           & 0.1666            &      (23)         &    (a) \\
                       &   UU Aqr          &      & 0.16381          & 0.1751            &      (29)         &        \\
                       &   TV Col          &      & 0.22860          & 0.264             &      (30)         &    (a) \\
                
\hline
\hline
                        
\end{tabular}

 \begin{tablenotes}

\item
References:
            
(1) \cite{pattersonwarner98}, (2) \cite{Mason2013CPpup}, (3) \citet{Olech2002V1974Cyg}, (4) \citet{Shugarov2002V1974Cyg}, (5) \cite{woudt2001}, (6) \cite{Mroz2015}, (7) \cite{Lipkin2008V4633Sgr}, 
(8) \cite{patterson97V606Aql}, (9) \cite{Peters2006}, (10) \cite{woudt2002}, (11) \cite{linda2008}, 
(12) This work, (13) \cite{Kato2013SUumaSH}, (14) \cite{patterson2013}, (15) \cite{Rolfe2000V348PupSH}, (16) \cite{Dai2010},  (17) \cite{Shafter90V795HerNL}, (18) \cite{Simon2012V795HerNL}, (19) \cite{Williams2010}, (20) \cite{kato2013NL}, (21) \cite{taylor98V592Cass}, (22) \cite{Rude2012LQPeg}, (23) \cite{patterson99}, (24) \cite{skillman95MVLyrNL}, (25) \cite{Stanishev2004DWUrsNL}, (26) \cite{Stanishev2001TTari}, (27) \cite{patterson2001}, (28) \cite{kato2001BZCam}, (29) \cite{patterson2005}, (30) \cite{retter2003TVCol}\\

Remarks:

(a) shows also negative superhumps \\
(b) possibly asynchronous rotation of the white dwarf (see more detail in the text)  \\ 
(c) possibly the remnant of a classical nova, observed on 101 December 30 \citep{patterson2013}\\

\end{tablenotes}  
\end{threeparttable}
\label{tab:t2}
\end{table*}

\subsection{RR Pic in context with other superhumpers}

Among all CVs which are neither dwarf novae nor AM CVn stars, positive superhumps (with $\epsilon$ $>$0) have been reported for seven classical post-novae and 17 nova-like variables (Table~\ref{tab:t2}). RR Pic is the sole Nb type nova, while all the remaining six cases belong to the Na class with a more rapid decline rate. This relation is similar to the general ratio between Na and Nb novae, 83\% and 17\% respectively, according to The International Variable Star index \citep[VSX,][]{watsonVSX}.

\cite{dobrotka2008} collected a sample of those post-novae for which the mass ratio $q=M_{2}/M_{1}$ could be derived, in order to correlate $q$ with the superhump occurrence, which is expected theoretically for cases with $q \leq 0.35$. In their sample (see their table 2) a total of 21 post-novae have mass ratio below or near to the critical value $q=0.35$, but actually only seven of those, i.e. a third, have been reported to show superhumps. On the other hand, only in two of eleven data sets of RR Pic superhumps were present (18\%). Therefore, it is possible that all post-novae with $q \leq 0.35$ occasionally develop superhumps, but that these events escape detection because of incomplete monitoring. 

For RR Pic, \cite{haefner82} had presented a model with a ratio $q= 0.42$, larger than the critical value expected to present superhumps. While their adopted WD mass of $0.95\,M_{\sun}$ is in good agreement with a recent model by \cite{mass1rrpic2017} based on far UV spectroscopy ($\sim 1\,M_{\sun}$) the mass of the secondary seems to be smaller, according to a systematic investigation the properties of the secondary stars in CVs by \cite{smithdhillon98M2}. When applying the empirical orbital period-mass relation given by these authors we get $q=0.33\pm 0.08$ for RR Pic (according to their equation (9)), well in agreement with \cite{linda2008}, who determined $q=0.31$ from the period excess relation $\epsilon=0.0860$, and the mass ratio relation formulated by \cite{patterson2005} valid for high $q$ ($\epsilon=0.18q+0.29q^{2}$).

\begin{figure}
  \centering  
  \resizebox{0.45\textwidth}{!}{\includegraphics{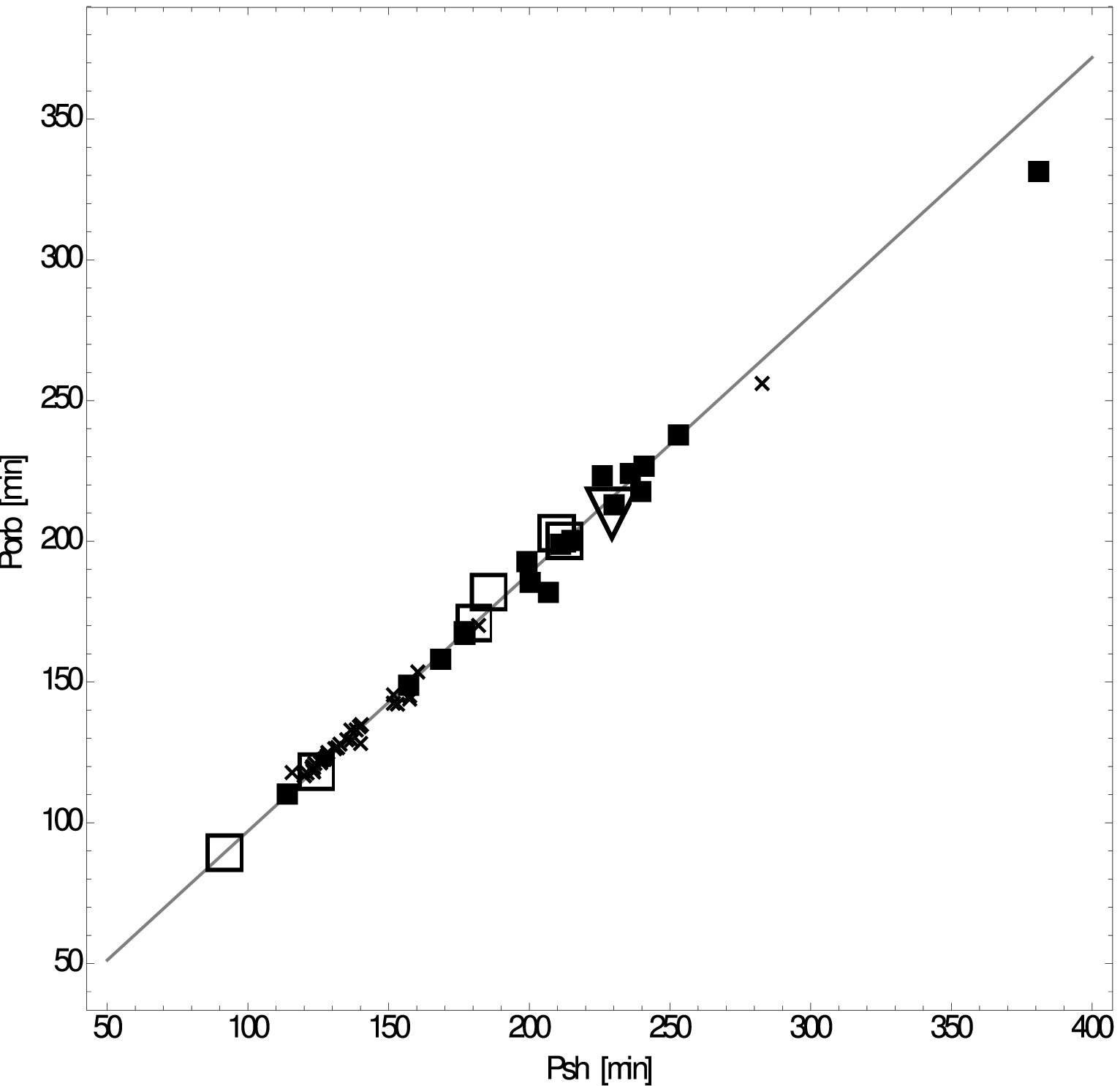}}
 \caption{The Stolz--Schoembs relation ($P_\mathrm{sh}-P_\mathrm{orb}$) for all sub-types of CVs in the entire range of $P_\mathrm{sh}$. Dwarf novae with $P_\mathrm{sh} > 120$\,min are represented as crosses, all nova-like stars as filled squares, all classical novae as empty squares and the classical nova RR Pic as an inverted empty triangle. The grey line corresponds to the linear fit found by \protect\cite{gaensicke2009} for SU UMa type dwarf novae with $P_\mathrm{sh} < 120$ min.}
  \label{fig:fitPs}
\end{figure}

We also investigated the role of RR Pic in context of the well-known close relation between $P_\mathrm{orb}$ and $P_\mathrm{sh}$ found first by \cite{stolzShoembs84}. \cite{gaensicke2009} gave a new version of this relation, in the form

\begin{equation}
P_\rmn{orb}= a + b\cdot P_\rmn{sh}
\label{eq:fitP}
\end{equation}
for the period range $P_\mathrm{sh}$ < 120 minutes, i.e. below the period gap between 2 and 3 hours. Most SU UMa stars are located below this gap, and the Stolz--Schoembs relation is now a popular tool for period determination because it is normally much easier to observe superhumps than any orbital variation, at least for non-eclipsing cases. \cite{gaensicke2009} derived a standard deviation of only 0.53 minutes for this method to get $P_\mathrm{orb}$ from equation~(\ref{eq:fitP}). We have repeated the calculation of these authors including new data, and extended it for cases with $P_\mathrm{sh} > 2$\,h. The results are listed in Table~\ref{tab:t3}. For the dwarf novae below the period gap, slope $a$ and standard deviation ($\sigma=0.89$ minutes) are larger in the actually available sample of 182 cases, compared to \cite{gaensicke2009}. If we consider all dwarf novae, we even get $\sigma=1.32$ minutes. Nova-like stars seem to have smaller slopes than the remaining groups while the post-novae behave similar to the dwarf novae, however suffering from the small number of only seven post-novae. All groups display larger standard deviations above the period gap, compared to those of $P_\mathrm{orb} < 2$\,h. Generally, the differences between the groups considered here are marginal, confirming that the Stolz--Schoembs relation seems to be valid for most superhumpers of any sub-type of CVs, as shown in Fig. \ref{fig:fitPs}, where all novae and nova-like stars are individualised, as well as the SU UMa type dwarf novae above the period gap ($P_\mathrm{orb} > 3$\,h). 

\begin{table}
\begin{minipage}{1.0\columnwidth}

 \caption{$P_\mathrm{sh}-P_\mathrm{orb}$ linear fit considering different sub-samples of CVs.  $a$ and $b$ correspond to the parameters of the linear fit, $\rmn{\sigma}_{\rmn{a}}$ and  $\rmn{\sigma}_{\rmn{b}}$ to the mean errors,  $\rmn{\sigma}$ refers to the standard deviation. $N$ represents the number of stars in each sub-sample. The first row shows the fit found by \protect\cite{gaensicke2009}.}

\setlength{\tabcolsep}{0.16cm}
\begin{tabular}{llllllr}

\hline
\hline
                       &\multicolumn{4}{c}{ $P_\mathrm{orb}~(\rmn{min})= a + b\cdot P_\mathrm{sh}~(\rmn{min})$ }& &\\
                       &$a$&$\rmn{\sigma}_{\rmn{a}}$ &$b$    &$\rmn{\sigma}_{\rmn{b}}$ & $\rmn{\sigma}$         &   $N$ \\

\hline
\hline
\cite{gaensicke2009}   &         &              &         &                &          &     \\
DN~~P $\leq$ 120\,min  & 5.39    & 0.52         &  0.9162 &  0.0052        &  0.53    &  68 \\
\hline
DN~~P $\leq$ 120\,min  &  3.66   & 0.49         &  0.9329 &  0.0053        &  0.89    & 182 \\
\hline
DN~~P $>$ 120\,min     &  14.09  & 1.62         &  0.8474 &  0.0114        &  1.90    & 35  \\
\hline
All DN                 &  7.09   & 0.40         &  0.8964 &  0.0038        &  1.32    & 217 \\
\hline
NL                     & 17.77   & 6.14         &  0.8418 &  0.0280        &  6.26    & 17  \\
\hline
Na, Nb                 &  6.11   & 6.36         &  0.9193 &  0.0352        &  4.33    & 7   \\
\hline
NL\,$+$ (Na,Nb)        &  16.04  & 4.62         &  0.8531 &  0.0221        &  5.96    & 24  \\
\hline
All CVs                &  7.83   & 0.42         &  0.8899 &  0.0036        &  2.32    & 241 \\

\hline
  \hline
\end{tabular}
\label{tab:t3}
\end{minipage}

\end{table}

Of special interest is the case of V4633~Sgr, because \cite{Lipkin2008V4633Sgr} have suggested that the second periodicity could correspond to the spin of the magnetic white dwarf of this system that rotates nearly synchronously with the orbital revolution. However, it would be a very strange coincidence that the white dwarf rotation fits quite well the Stolz--Schoembs relation; we therefore believe that we could have a superhump also in this case. Finally, for 29\% of the classical novae and 41\% of the nova-likes stars also negative superhumps ($P_\mathrm{sh} < P_\mathrm{orb}$) have been detected, as marked in Table \ref{tab:t2}. The simultaneous presence of superhumps above and below the orbital period strongly suggests that independent physical mechanisms are responsible for this behaviour. The longer period (``positive superhump'') could arise from the prograde motion of the line of apsides of an eccentric disc \citep{vogt82a} while the shorter period (``negative superhump'') could be due to the retrograde motion of the line of nodes of a tilted disc \citep{skillman98}.

\section{Conclusions}

We present the first systematic analysis of collected optical light curves of the classical nova RR Pic spanning 42 years, from 1972 to 2014, including new observations 2013--2014 obtained by one us (FJH) with a robotic telescope in San Pedro de Atacama. We found that the light curve shape of the orbital hump and its amplitude varies significantly over the different observing epochs, suggesting that the accretion disc is undergoing long-term changes. Furthermore, we corroborate the superhump detection made by \citet{linda2008} and additionally we find evidence for the presence of most likely the same positive superhump in 2007 CBA data, a period $\sim$ 9\% longer than the orbital period. Equally important was the fact that CBA data have been obtained in 2005 only one month before those by \citet{linda2008}, but did not reveal any evidence for superhumps, implying the superhump is a repetitive, but sporadic event, arising quickly. 

We give a list of seven post-novae and 17 nova-like stars with positive superhumps recorded, and conclude that in order to determine whether the superhump phenomenon is an unusual event or common in classical novae, it is imperative to monitor post-novae systematically for long time intervals. 
 In addition, we also determine a revised version of the Stolz--Schoembs relation between $P_\mathrm{sh}$ and $P_\mathrm{orb}$ for dwarf novae, classical novae and nova-like stars, and conclude that this relation is of general validity for all superhumpers among CVs.


This apparently also includes post-novae, with RR Pic being just one more case in a still small sub-sample. This should not come as a surprise, since it has long been suspected that the nova eruption represents a recurrent event in a CV long-term cycle \citep{vogt82b, shara86}, with more recent observational evidence corroborating this hypothesis \citep[e.g.][]{patterson2013, mroz2016, shara2017}. The fact that post-novae present the same phenomenon as other CVs, further strengthens our understanding of the long-term evolution of CVs.

\section*{Acknowledgments}
CT, NV and MV acknowledge financial support from FONDECYT Regular No.1170566. IFM thanks to CONICYT-PFCHA/Doctorado Nacional/2017-21171099 by doctoral fellowship and to European Southern Observatory (ESO) for offering stay in their offices in the city of Santiago. IFM, NV, CT and MV acknowledge support by the Centro de Astrof\'isica de Valpara\'iso (CAV).\\
All authors give special thanks to Joe Patterson and Jonathan Kemp by send us the data available from CBA database.




\bibliographystyle{mnras}
\bibliography{ref} 




\appendix




\bsp	
\label{lastpage}
\end{document}